\pdfoutput=1  
\documentclass[conference,a4paper]{IEEEtran}
\IEEEoverridecommandlockouts
\usepackage{cite}
\usepackage{amsmath,amssymb,amsfonts}
\usepackage{graphicx}
\usepackage{textcomp}
\usepackage{xcolor}
\usepackage{booktabs}
\usepackage{subcaption}
\usepackage{xfrac}
\usepackage{nicematrix}
\usepackage{tikz}
\usepackage{pgfplots}
\pgfplotsset{compat=1.18}
\usetikzlibrary{arrows.meta}
\usetikzlibrary{patterns}
\definecolor{cmapBlue}{rgb}{0.121569,0.466667,0.705882}
\definecolor{cmapOrange}{rgb}{1.000000,0.498039,0.054902}
\definecolor{cmapGreen}{rgb}{0.172549,0.627451,0.172549}
\definecolor{cmapRed}{rgb}{0.839216,0.152941,0.156863}
\definecolor{cmapPurple}{rgb}{0.580392,0.403922,0.741176}
\definecolor{cmapBrown}{rgb}{0.549020,0.337255,0.294118}
\definecolor{cmapPink}{rgb}{0.890196,0.466667,0.760784}
\definecolor{cmapGray}{rgb}{0.498039,0.498039,0.498039}
\definecolor{cmapOlive}{rgb}{0.737255,0.741176,0.133333}
\definecolor{cmapCyan}{rgb}{0.090196,0.745098,0.811765}
\definecolor{cmapIndigo}{rgb}{0.227451,0.003922,0.513725}
\definecolor{cmapDarkgreen}{rgb}{0.000000,0.262745,0.003922}
\definecolor{cmapMint}{rgb}{0.058824,1.000000,0.662745}
\definecolor{cmapCrimson}{rgb}{0.368627,0.000000,0.250980}
\definecolor{cmapLavender}{rgb}{0.737255,0.737255,1.000000}
\definecolor{cmapSand}{rgb}{0.847059,0.686275,0.635294}

\definecolor{cmapBlueDark}{rgb}{0.097255,0.373334,0.564706}       
\definecolor{cmapOrangeDark}{rgb}{0.800000,0.398431,0.043922}     
\definecolor{cmapGreenDark}{rgb}{0.138039,0.501961,0.138039}      
\definecolor{cmapRedDark}{rgb}{0.671373,0.122353,0.125490}        
\definecolor{cmapPurpleDark}{rgb}{0.464314,0.323138,0.592941}     
\definecolor{cmapBrownDark}{rgb}{0.439216,0.269804,0.235294}      
\definecolor{cmapPinkDark}{rgb}{0.712157,0.373334,0.608627}       
\definecolor{cmapGrayDark}{rgb}{0.398431,0.398431,0.398431}       
\definecolor{cmapOliveDark}{rgb}{0.589804,0.592941,0.106666}      
\definecolor{cmapCyanDark}{rgb}{0.072157,0.596078,0.649412}       
\definecolor{cmapIndigoDark}{rgb}{0.181961,0.003138,0.410980}     
\definecolor{cmapDarkgreenDark}{rgb}{0.000000,0.210196,0.003138}  
\definecolor{cmapMintDark}{rgb}{0.047059,0.800000,0.530196}       
\definecolor{cmapCrimsonDark}{rgb}{0.294902,0.000000,0.200784}    
\definecolor{cmapLavenderDark}{rgb}{0.589804,0.589804,0.800000}   
\definecolor{cmapSandDark}{rgb}{0.677647,0.549020,0.508235}       

\definecolor{cmapBlueDarker}{rgb}{0.072941,0.280000,0.423529}     
\definecolor{cmapOrangeDarker}{rgb}{0.600000,0.298823,0.032941}   
\definecolor{cmapGreenDarker}{rgb}{0.103529,0.376471,0.103529}    
\definecolor{cmapRedDarker}{rgb}{0.503530,0.091765,0.094118}      
\definecolor{cmapPurpleDarker}{rgb}{0.348235,0.242353,0.444706}   
\definecolor{cmapBrownDarker}{rgb}{0.329412,0.202353,0.176471}    
\definecolor{cmapPinkDarker}{rgb}{0.534118,0.280000,0.456470}     
\definecolor{cmapGrayDarker}{rgb}{0.298823,0.298823,0.298823}     
\definecolor{cmapOliveDarker}{rgb}{0.442353,0.444706,0.080000}    
\definecolor{cmapCyanDarker}{rgb}{0.054118,0.447059,0.487059}     
\definecolor{cmapIndigoDarker}{rgb}{0.136471,0.002353,0.308235}   
\definecolor{cmapDarkgreenDarker}{rgb}{0.000000,0.157647,0.002353} 
\definecolor{cmapMintDarker}{rgb}{0.035294,0.600000,0.397647}     
\definecolor{cmapCrimsonDarker}{rgb}{0.221176,0.000000,0.150588}  
\definecolor{cmapLavenderDarker}{rgb}{0.442353,0.442353,0.600000} 
\definecolor{cmapSandDarker}{rgb}{0.508235,0.411765,0.381176}     

\definecolor{cmapBlueDarkest}{rgb}{0.048628,0.186667,0.282353}    
\definecolor{cmapOrangeDarkest}{rgb}{0.400000,0.199216,0.021961}  
\definecolor{cmapGreenDarkest}{rgb}{0.069020,0.250980,0.069020}   
\definecolor{cmapRedDarkest}{rgb}{0.335686,0.061176,0.062745}     
\definecolor{cmapPurpleDarkest}{rgb}{0.232157,0.161569,0.296470}  
\definecolor{cmapBrownDarkest}{rgb}{0.219608,0.134902,0.117647}   
\definecolor{cmapPinkDarkest}{rgb}{0.356078,0.186667,0.304314}    
\definecolor{cmapGrayDarkest}{rgb}{0.199216,0.199216,0.199216}    
\definecolor{cmapOliveDarkest}{rgb}{0.294902,0.296470,0.053333}   
\definecolor{cmapCyanDarkest}{rgb}{0.036078,0.298039,0.324706}    
\definecolor{cmapIndigoDarkest}{rgb}{0.090980,0.001569,0.205490}  
\definecolor{cmapDarkgreenDarkest}{rgb}{0.000000,0.105098,0.001569} 
\definecolor{cmapMintDarkest}{rgb}{0.023530,0.400000,0.265098}    
\definecolor{cmapCrimsonDarkest}{rgb}{0.147451,0.000000,0.100392} 
\definecolor{cmapLavenderDarkest}{rgb}{0.294902,0.294902,0.400000} 
\definecolor{cmapSandDarkest}{rgb}{0.338824,0.274510,0.254118}    

\begin{document}

\title{Characterization of Helicopter Rotor Blade Modulation in UHF and Microwave Bands\\{\Large Extended Version}\thanks{This is an extended version of a paper accepted for presentation at the IEEE Ukrainian Microwave Week (UkrMW), 2026.}}

\author{
\IEEEauthorblockN{Wilhelm Keusgen}
\IEEEauthorblockA{\textit{Technische Universit\"at Berlin} \\
Berlin, Germany \\
ORCID 0000-0002-2335-8270}
\and
\IEEEauthorblockN{Alper Schultze, Mathis Schmieder, Michael Peter}
\IEEEauthorblockA{\textit{Fraunhofer Heinrich-Hertz-Institut, HHI} \\
Berlin, Germany\\
forename.surname@hhi.fraunhofer.com}
\and
\IEEEauthorblockN{Taro Eichler, Friedrich Lipp}
\IEEEauthorblockA{\textit{Rohde \& Schwarz GmbH \& Co. KG} \\
M\"unchen, Germany\\
forename.surname@rohde-schwarz.com}
}

\maketitle

\begin{abstract}
This paper presents metrics and measurement methods for the experimental characterization of helicopter rotor blade modulation of the wireless propagation channel. Broadband bistatic channel measurements were conducted at three widely separated carrier frequencies -- 302~MHz (UHF), 4.9~GHz (C-band), and 27.1~GHz (Ka-band) -- for three military helicopters (CH-53, Tiger, and UMAT) at the Bundeswehr Technical and Airworthiness Center for Aircraft in Manching, Germany. The time-variant channel was analyzed in terms of the Doppler Power Spectrum and the RMS Doppler spread, the Doppler spectrogram, and the Power Time Profile. The RMS Doppler spread values were found to be moderate (up to 106~Hz), which is attributable to the predominantly lateral blade motion relative to the line-of-sight path. The Power Time Profiles revealed periodic blockage attenuations of up to 15~dB at Ka-band, increasing significantly with the carrier frequency; at 302~MHz, the blockage attenuation does not exceed 2.3~dB and the RMS Doppler spread stays below 10~Hz. The measured blockage attenuation was modeled using the double knife-edge diffraction model, showing good agreement with the measurements. The distinct micro-Doppler signatures visible in the spectrograms are identified as a potential resource for helicopter classification and passive radar applications.
\end{abstract}


\section{Introduction}
Next-generation secure tactical communication systems are expected to offer higher data rates, improved energy efficiency, global coverage, and connectivity, as well as extremely high reliability and low latency. This requires novel connected platforms for land, air, sea, and space, including the extension of the frequency range into the microwave bands. Helicopters are particularly challenging platforms in this context: rotor blade modulation produces periodic blockage attenuation of several dB to tens of dB and generates micro-Doppler sidebands at multiples of the blade-passage frequency, which degrade link reliability and complicate channel estimation \cite{kadar}, \cite{lemoscid}.

This paper investigates the impact of helicopter rotor blade modulation on the wireless propagation channel, with the helicopter-to-aircraft and helicopter-to-satellite links being prominent examples. By examining the time-variant broadband channel, we analyze the Doppler and micro-Doppler signatures produced by the rotating blades. Our study also focuses on the estimation and modeling of the signal attenuation caused by these dynamic effects.

Electromagnetic scattering by rotor blades and the resulting signal modulation have been studied analytically \cite{kadar}, numerically \cite{polycarpou}, \cite{pouliguen}, and experimentally \cite{desantis} since the early 1970s. A wideband characterization of the satellite-to-helicopter channel at Ku-band, measured with a sliding-correlator sounder and real helicopters, was reported in \cite{lemoscid}, identifying line-of-sight (LOS) blockage and blade-edge multipath as the two main propagation impairments. Airframe shadowing caused by fixed-wing aircraft has been studied at L- and C-band, showing attenuations of up to 47~dB lasting tens of seconds \cite{matolak}, and GNSS-based opportunistic measurements of the satellite-to-aircraft channel at L-band have been reported in \cite{jost}. In contrast to these prior works, the present study applies a broadband time-domain channel sounder across three widely separated frequency bands in a bistatic ground-based geometry to three different military helicopters.

The specific contributions of this paper are: (i) a set of metrics and measurement methods, adapted from the Bello framework for time-variant channels \cite{bello}, for the characterization of rotor blade modulation, comprising the Doppler Power Spectrum, the Doppler spectrogram, and the Power Time Profile; (ii) broadband bistatic measurement results for three military helicopters across three frequency bands; (iii) the quantification of the blade blockage attenuation as a function of carrier frequency and helicopter type; and (iv) the validation of the double knife-edge model for the prediction of blade blockage.

\section{Description of the Time-Variant Radio Channel}
The rotating blades introduce a strictly periodic time variance into the radio propagation channel, which makes the system-function framework of Bello \cite{bello} particularly well suited for its description: the Delay-Doppler Spread Function directly captures the blade-induced Doppler shifts, and the Doppler Power Spectrum naturally reveals the blade-passage periodicity. In contrast to most of the prior works, which are largely based on continuous-wave (CW) measurements or quasi-stationary analyses (e.g., \cite{polycarpou}, \cite{desantis}, \cite{kadar}), we therefore make use of this well-known framework for broadband time-variant radio channels.

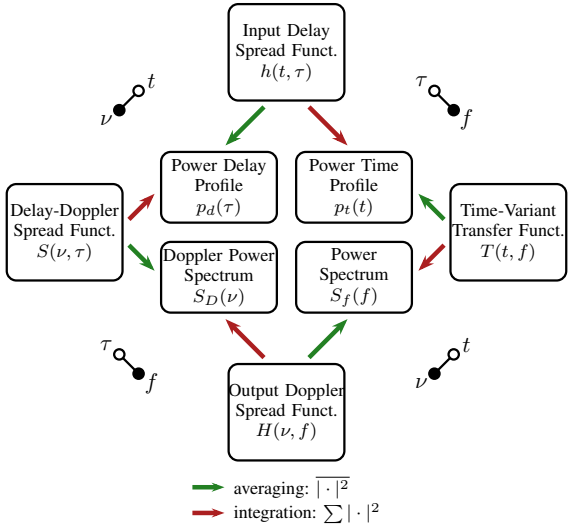
\begin{figure}[htbp]
  \centering
  \resizebox{0.85\columnwidth}{!}{\begin{tikzpicture}[x=1mm, y=1mm, line width=1pt,
  %
  %
  boxlabel/.style={align=center, font=\footnotesize, inner sep=0pt},
  %
  box/.style={rounded corners=1.5mm},
]



\draw[box] (-9, 20.5) rectangle (9, 35.5);
\node[boxlabel] at (0, 28)
  {Input Delay\\ Spread Funct.\\ $h(t,\tau)$};

\draw[box] (-9,-35.5) rectangle (9,-20.5);
\node[boxlabel] at (0,-28)
  {Output Doppler\\ Spread Funct.\\ $H(\nu,f)$};

\draw[box] (-43.5,-7.5) rectangle (-25.5, 7.5);
\node[boxlabel] at (-34.5, 0)
  {Delay-Doppler\\ Spread Funct.\\ $S(\nu,\tau)$};

\draw[box] (25.5,-7.5) rectangle (43.5, 7.5);
\node[boxlabel] at (34.5, 0)
  {Time-Variant\\ Transfer Funct.\\ $T(t,f)$};


\draw[box] (-19.5, 1.5) rectangle (-1.5, 12.5);
\node[boxlabel] at (-10.5, 7.0)
  {Power Delay\\ Profile\\ $p_d(\tau)$};

\draw[box] (1.5, 1.5) rectangle (19.5, 12.5);
\node[boxlabel] at (10.5, 7.0)
  {Power Time\\ Profile\\ $p_t(t)$};

\draw[box] (-19.5,-12.5) rectangle (-1.5,-1.5);
\node[boxlabel] at (-10.5,-7.0)
  {Doppler Power\\ Spectrum\\ $S_D(\nu)$};

\draw[box] (1.5,-12.5) rectangle (19.5,-1.5);
\node[boxlabel] at (10.5,-7.0)
  {Power\\ Spectrum\\ $S_f(f)$};

%
%

\tikzset{
  avgconn/.style={-{Stealth[length=2.5mm, width=1.8mm]},
                  line width=2pt, cmapGreenDark},
  sumconn/.style={-{Stealth[length=2.5mm, width=1.8mm]},
                  line width=2pt, cmapRedDark},
}

\draw[avgconn] ( -3.5, 19.5) -- ( -9.5, 13.5);  
\draw[sumconn] (  3.5, 19.5) -- (  9.5, 13.5);  

\draw[sumconn] (-24.5,  2.0) -- (-20.5,  6.0);  
\draw[avgconn] (-24.5, -2.0) -- (-20.5, -6.0);  

\draw[avgconn] ( 24.5,  2.0) -- ( 20.5,  6.0);  
\draw[sumconn] ( 24.5, -2.0) -- ( 20.5, -6.0);  

\draw[sumconn] ( -3.5,-19.5) -- ( -9.5,-13.5);  
\draw[avgconn] (  3.5,-19.5) -- (  9.5,-13.5);  

%

\node[anchor=north, inner sep=0pt]
  at (0, -37.5) {%
    \begin{tikzpicture}[x=1mm, y=1mm, baseline=0]
      \draw[cmapGreenDark, line width=2pt,
            -{Stealth[length=2mm, width=1.4mm]}]
        (0, 4) -- (5, 4);
      \node[anchor=west, font=\footnotesize, inner sep=1pt]
        at (6, 4) {averaging: $\overline{|\cdot|^2}$};
      \draw[cmapRedDark, line width=2pt,
            -{Stealth[length=2mm, width=1.4mm]}]
        (0, 0) -- (5, 0);
      \node[anchor=west, font=\footnotesize, inner sep=1pt]
        at (6, 0) {integration: $\textstyle\sum|\cdot|^2$};
    \end{tikzpicture}%
  };

%
%
%

\tikzset{
  opencircle/.style={draw, circle, minimum size=1.6mm,
                     inner sep=0pt, line width=1pt, fill=white},
  filledcircle/.style={draw, circle, minimum size=1.6mm,
                       inner sep=0pt, line width=1pt, fill=black},
  ftlabel/.style={font=\normalsize, inner sep=0pt},
}

\draw[line width=1pt] (-23, 22) -- (-26, 19);
\node[opencircle]   at (-23, 22) {};
\node[filledcircle] at (-26, 19) {};
\node[ftlabel] at (-21, 23.5) {$t$};
\node[ftlabel] at (-28, 17.5) {$\nu$};

\draw[line width=1pt] ( 23, 22) -- ( 26, 19);
\node[opencircle]   at ( 23, 22) {};
\node[filledcircle] at ( 26, 19) {};
\node[ftlabel] at ( 21, 23.5) {$\tau$};
\node[ftlabel] at ( 28, 17.5) {$f$};

\draw[line width=1pt] (-26,-19) -- (-23,-22);
\node[opencircle]   at (-26,-19) {};
\node[filledcircle] at (-23,-22) {};
\node[ftlabel] at (-28,-17.5) {$\tau$};
\node[ftlabel] at (-21,-23.5) {$f$};

\draw[line width=1pt] ( 26,-19) -- ( 23,-22);
\node[opencircle]   at ( 26,-19) {};
\node[filledcircle] at ( 23,-22) {};
\node[ftlabel] at ( 28,-17.5) {$t$};
\node[ftlabel] at ( 21,-23.5) {$\nu$};

\end{tikzpicture}}
  \caption{System functions of the time-variant radio channel and the corresponding one-dimensional power density functions.}
  \label{fig:bello}
\end{figure}

The time-variant radio channel can be fully described by four system functions: the Time-Variant Transfer Function, the Input Delay Spread Function, the Delay-Doppler Spread Function, and the Output Doppler Spread Function \cite{bello} (see Fig.~\ref{fig:bello}). Each of the system functions depends on two of the four variables time $t$, delay $\tau$, frequency $f$, and Doppler frequency $\nu$; the pairs $\tau$,~$f$ and $t$,~$\nu$ are conjugate variables related by the Fourier transform ($\circ\!\!-\!\!\bullet$) or its inverse ($\bullet\!\!-\!\!\circ$), respectively. From the system functions, four real-valued one-dimensional power density functions, known as profiles or spectra, can be derived by eliminating one of the two independent variables. This elimination is implemented either by averaging the powers (squared magnitudes) in the respective dimension ($\overline{P}$) or by the summation of the powers ($\Sigma P$), as indicated in Fig.~\ref{fig:bello}. Further details on the derivation of the one-dimensional power density functions can be found in \cite{keusgen}.
\section{Helicopters and Measurement Setup}
Three military helicopters were investigated during the study: the CH-53 from Sikorsky Aircraft Corporation (CH-53), the Eurocopter Tiger from Airbus Helicopters (Tiger), and the experimental UAV R-350 from UMS Skeldar AG (UMAT). The measurements took place at the Bundeswehr Technical and Airworthiness Center for Aircraft in Manching, Germany. The influence of the main rotors and of the tail rotors was investigated by means of broadband bistatic propagation measurements at three different frequencies within the UHF, C-, and Ka-bands (UMAT: C- and Ka-band only). In this paper, we report the results for the main rotors; the tail rotor results are left for a future publication. Additional static measurements with a stopped rotor were performed to verify the measurement setup.

The measurements were performed with the transmitter (Tx) located on the ground close to the helicopter and the receiver (Rx) positioned on an airport tower at a height of 40~m, creating a slanted LOS path that was obstructed by the rotor blades. Fig.~\ref{fig:geometry}(a) shows a side view of the measurement geometry, and Fig.~\ref{fig:geometry}(b) gives an impression of the setup for the CH-53, showing the transmitter in the foreground. Depending on the helicopter position, the LOS distance was between approximately 80~m and 95~m; the elevation angle of the LOS path was approximately 25\textdegree{} for the Tiger and 30\textdegree{} for the CH-53 and the UMAT. In the realized setup, the LOS path was obstructed by the outer third of the rotor blades; thus, the relevant blade speeds were around 70\% of the tip speeds listed in Table~\ref{tab:helicopter_parameters}. The theoretical maximum Doppler frequencies in the table, which were calculated for a reflecting blade tip, are reduced accordingly in practice. Table~\ref{tab:geometry} summarizes the resulting per-helicopter geometry: the LOS elevation angle, the LOS distance $d$, and the distance $d_1$ from the Tx to the point where the LOS path crosses the rotor disk (blade intercept point).
\begin{figure}[!t]
  \centering
  \begin{subfigure}{\columnwidth}
    \centering
    \resizebox{\columnwidth}{!}{\begin{tikzpicture}[
  x=1mm, y=1mm,
  line width=0.5pt, line cap=round, line join=round,
  dimline/.style={
    {Stealth[length=1.6mm, width=1.2mm]}-{Stealth[length=1.6mm, width=1.2mm]},
    line width=0.4pt
  },
  thinline/.style={line width=0.3pt},
]

  \draw[line width=1pt] (0, 0) -- (120, 0);

  \draw (10, 0) -- (10, 40) -- (18, 40) -- (18, 0);

  \draw (10, 28) -- (8, 28) -- (8, 40) -- (10, 40);    
  \draw (18, 28) -- (20, 28) -- (20, 40) -- (18, 40);  
  \draw (8, 32) -- (10, 32);
  \draw (8, 36) -- (10, 36);
  \draw (18, 32) -- (20, 32);
  \draw (18, 36) -- (20, 36);

  \draw[thinline] (4, 40) -- (8, 40);
  \draw[thinline] (20, -9) -- (20, 0);
  \draw[thinline] (105, -9) -- (105, 0);

  \draw[dimline] (6, 0) -- (6, 40);
  \node[anchor=east, inner sep=2pt] at (6, 20) {40\,m};

  \draw[dimline] (20, -7) -- (105, -7);
  \node[anchor=north, inner sep=2pt] at (62.5, -7) {85\,m};

  \node[anchor=north, inner sep=2pt] at (14, -2) {Tower};

  \begin{scope}[shift={(92.57, 0.733)}, scale=0.15323, line cap=butt,
                color=black, line width=0.7pt]
    \draw (-42.420, 19.453) -- (-0.062, 19.430) -- (42.420, 19.454);
    \draw (-0.062, 19.430) -- (-0.072, 17.003);
    \draw (-1.954, 14.497)
       -- (-1.938, 16.742) -- (-0.072, 17.003)
       -- ( 1.746, 16.903) -- ( 1.876, 14.520)
       -- ( 5.094, 13.600) -- ( 7.188, 11.188)
       arc[start angle= 17.714, end angle=  -8.422, radius=14.791]
       -- ( 7.658,  2.111) -- ( 5.533, -0.661)
       -- (-5.612, -0.652) -- (-7.763,  2.148)
       -- (-7.825,  4.500)
       arc[start angle=189.256, end angle= 161.897, radius=14.049]
       -- (-4.858, 13.802)
       -- cycle;
    \draw (-1.954, 14.497) -- ( 1.876, 14.520);
    \draw (-7.825, 4.500)
       arc[start angle=-111.167, end angle=-86.906, radius=18.702]
       arc[start angle= -93.361, end angle=-68.335, radius=18.213];
    \draw (-0.042, 14.292) -- (-0.062, 3.265);
    \draw (-5.612, -0.652) -- (-10.110, -4.782);
    \draw ( 5.533, -0.661) -- ( 10.066, -4.785);
  \end{scope}

  \draw[
    -{Stealth[length=2.5mm, width=1.6mm]},
    line width=0.8pt,
    dashed,
    shorten >=1mm, shorten <=1mm
  ] (105, 0) -- (20, 40);

  \draw[line width=0.4pt]
    (100, 0) arc[start angle=180, end angle=154.8, radius=5];
  \draw[thinline] (100.7, 1.7) -- (102.8, 5);
  \node[fill=white, inner sep=1.5pt] at (102.8, 5) {$25^\circ$};

  \fill (20, 40) circle (0.7);
  \fill (105, 0) circle (0.7);

  \node[anchor=south west, inner sep=2pt] at (20.5, 40.5) {$R_x$};
  \node[anchor=south west, inner sep=2pt] at (105.5, 0.5) {$T_x$};

\end{tikzpicture}}
    \caption{}
    \label{fig:geometry_schematic}
  \end{subfigure}\par\smallskip
  \begin{subfigure}{\columnwidth}
    \centering
    \includegraphics[width = 0.92\columnwidth]{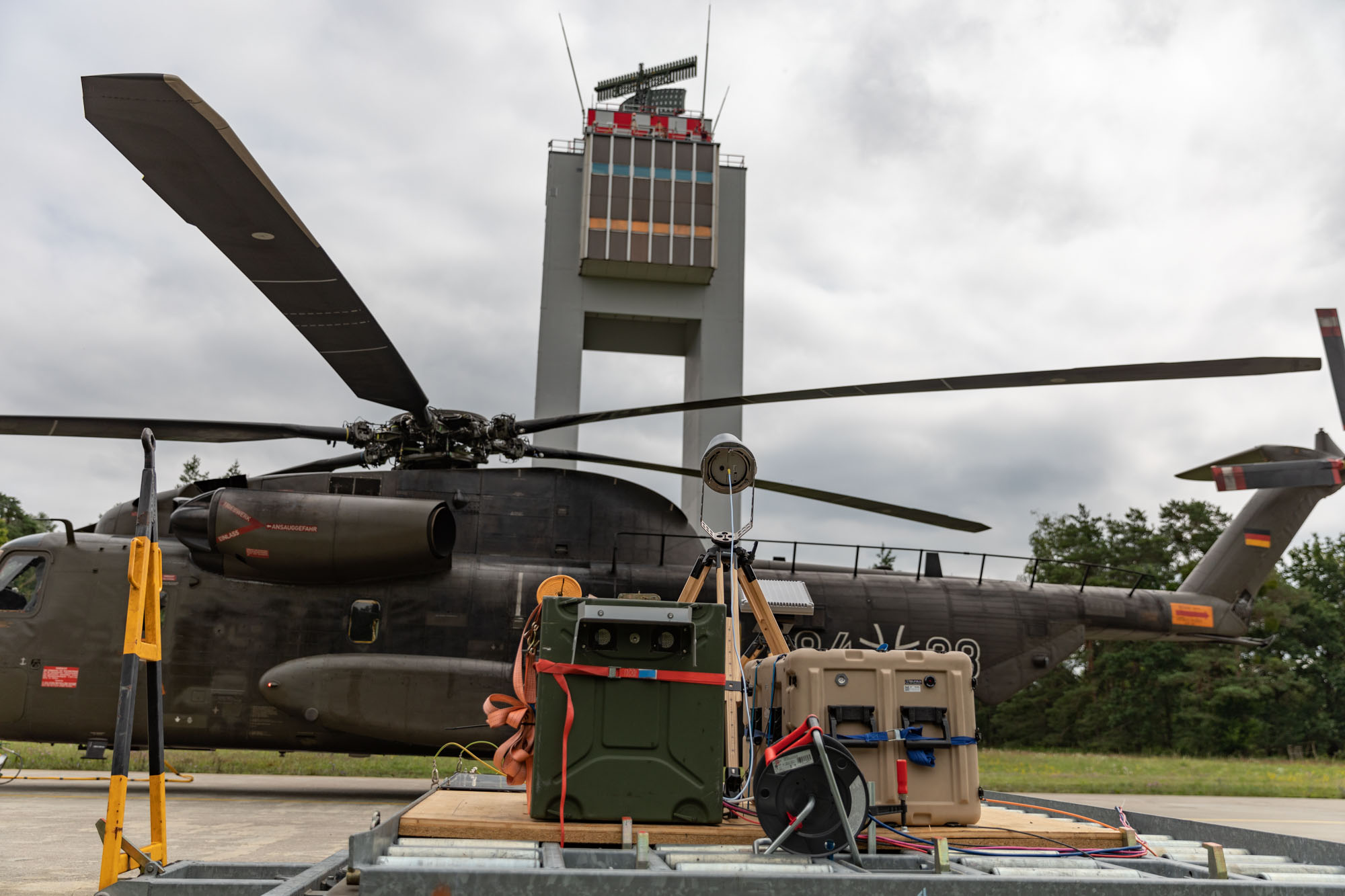}
    \caption{}
    \label{fig:geometry_photo}
  \end{subfigure}
  \caption{Measurement setup: (a) side view of the geometry, drawn to scale for the Tiger (LOS elevation angle 25\textdegree{}; the LOS path crosses the rotor disk at 70\% of the blade radius); (b) setup for the CH-53 with the transmitter in the foreground.}
  \label{fig:geometry}
\end{figure}
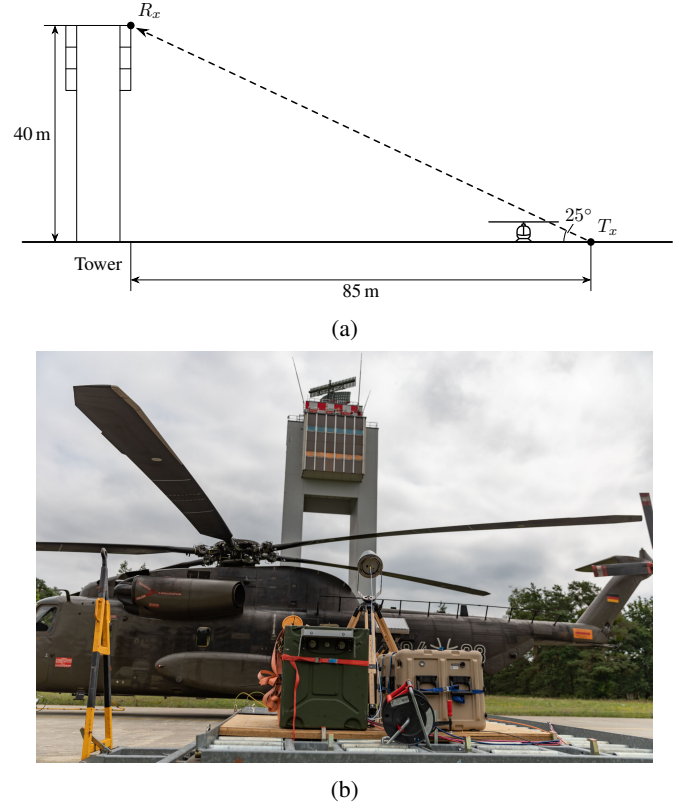
\begin{table}[htbp]
\caption{Rotor characteristics of the investigated helicopters}
\label{tab:helicopter_parameters}
    \begin{center}
        \begin{NiceTabular}[c]{l l l l}
        \CodeBefore
        \rowcolor{gray!50}{1-1}
       \rowcolors{2}{gray!15}{}
        \Body
            \toprule
			 & \textit{UMAT} & \textit{Tiger} & \textit{CH-53}\\
			\midrule
   No. of rotor blades ($N_\mathrm{b}$) & 3 & 4 & 6\\
   Rotor frequency ($f_\mathrm{rot}$) & 14.17 Hz & 5.45 Hz & 3.08 Hz\\
   Rotor radius (approx.) & 1.8 m & 6.5 m & 11.0 m\\
   Width of blades (approx.) & 0.3 m & 0.5 m & 0.7 m\\
   Blade-to-blade time & 23.5 ms & 45.9 ms & 54.1 ms\\
   Speed at blade tip & 159.3 \sfrac{m}{s} & 222.6 \sfrac{m}{s} & 213.2 \sfrac{m}{s}\\
   Max. Doppler freq. at 27.1 GHz & 28.7 kHz & 40.1 kHz & 38.4 kHz\\
			\bottomrule
        \end{NiceTabular}
    \end{center}
\end{table}
\begin{table}[htbp]
\caption{Measurement geometry for the individual helicopters}
\label{tab:geometry}
    \begin{center}
        \begin{NiceTabular}[c]{l l l l}
        \CodeBefore
        \rowcolor{gray!50}{1-1}
        \rowcolors{2}{gray!15}{}
        \Body
            \toprule
			 & \textit{UMAT} & \textit{Tiger} & \textit{CH-53}\\
			\midrule
   LOS elevation angle & 30.5\textdegree & 25.2\textdegree & 30.2\textdegree\\
   LOS distance ($d$) & 78.8 m & 93.9 m & 79.5 m\\
   Distance Tx to blade intercept ($d_1$) & 2.5 m & 8.8 m & 16.9 m\\
			\bottomrule
        \end{NiceTabular}
    \end{center}
\end{table}

The time-variant propagation channels were measured by means of a time-domain channel sounder. In this approach, the channel is conceptually probed with a train of short impulses, which directly yields a train of channel impulse responses at the receiver. This train corresponds to the Input Delay Spread Function $h(t,\tau)$ (see Fig.~\ref{fig:bello}), which serves as the starting point for the derivation of all other system functions. Practically, the train of impulses is created by the autocorrelation function of a perfect correlation sequence, with the correlation implemented at the receiver side. The length $M$ of the correlation sequence must be chosen appropriately to capture the maximum delays as well as the maximum Doppler frequencies. We propose the use of Zadoff--Chu sequences, which allow for the observation of time-variant channels even in the presence of a strong phase shift within one period of the correlation sequence \cite{wittig}; the achievable dynamic range of this sounder class is analyzed in \cite{wittig2020}. Further details about this channel sounding approach can be found in \cite{keusgen}.

\begin{figure}[htbp]
  \centering
  \begin{subfigure}{\columnwidth}
    \centering
    \resizebox{0.86\columnwidth}{!}{\begin{tikzpicture}[x=1mm, y=1mm, line width=1pt,
  %
  %
  boxlabel/.style={align=center, font=\footnotesize, inner sep=0pt},
  %
  arrlabel/.style={font=\footnotesize, inner sep=1pt},
  %
  box/.style={rounded corners=1.5mm},
  %
  conn/.style={-{Stealth[length=2.5mm, width=1.8mm]}, line width=1pt},
]



\draw[box] (-22, 19) rectangle (-2, 31);
\node[boxlabel] at (-12, 25)
  {Antenna\\10-118-37};

\draw[box] (2, 19) rectangle (22, 31);
\node[boxlabel] at (12, 25)
  {Antenna\\R\&S\textsuperscript{\textregistered}HL050};


\draw[box] (-42, -6) rectangle (-22, 6);
\node[boxlabel] at (-32, 0)
  {Power\\Amplifier};

\draw[box] (-10, -6) rectangle (10, 6);
\node[boxlabel] at (0, 0)
  {RF Switch\\R\&S\textsuperscript{\textregistered}OSP130};

\draw[box] (22, -6) rectangle (42, 6);
\node[boxlabel] at (32, 0)
  {Power\\Amplifier};


\draw[box] (-13, -30) rectangle (13, -18);
\node[boxlabel] at (0, -24)
  {Signal Generator\\R\&S\textsuperscript{\textregistered}SMW200A};

\draw[box] (-13, -50) rectangle (13, -38);
\node[boxlabel] at (0, -44)
  {Fraunhofer HHI\\Synchronomat};


%
%
\draw[conn] (-11,  3) -- (-21,  3);            
\draw[conn] (-21, -3) -- (-11, -3);            
\node[arrlabel, anchor=south] at (-16, 4) {5\,GHz};

\draw[conn] ( 11,  3) -- ( 21,  3);            
\draw[conn] ( 21, -3) -- ( 11, -3);            
\node[arrlabel, anchor=south] at ( 16, 4) {27\,GHz};

%
%
\draw       (  0,  7) -- (  0, 10);            
\draw       (-12, 10) -- ( 12, 10);            
\draw[conn] (-12, 10) -- (-12, 18);            
\draw[conn] ( 12, 10) -- ( 12, 18);            

\node[arrlabel, anchor=east]             at (-13, 14) {300\,MHz};
\node[arrlabel, anchor=west, align=left] at ( 13, 14) {5\,GHz\\27\,GHz};

%
%
\draw[conn] (0, -17) -- (0, -7);
\node[arrlabel, anchor=north west, align=left] at (1, -7.5)
  {300\,MHz\\5\,GHz\\27\,GHz};

%
%
\draw[conn] (0, -37) -- (0, -31);
\node[arrlabel, anchor=west] at (1, -34) {10\,MHz};

\end{tikzpicture}}
    \caption{}
    \label{fig:block_tx}
  \end{subfigure}\par\smallskip
  \begin{subfigure}{\columnwidth}
    \centering
    \resizebox{0.48\columnwidth}{!}{\begin{tikzpicture}[x=1mm, y=1mm, line width=1pt,
  %
  %
  boxlabel/.style={align=center, font=\footnotesize, inner sep=0pt},
  %
  arrlabel/.style={font=\footnotesize, inner sep=1pt},
  %
  box/.style={rounded corners=1.5mm},
  %
  conn/.style={-{Stealth[length=2.5mm, width=1.8mm]}, line width=1pt},
]



\draw[box] (-22, -3) rectangle (-2, 9);
\node[boxlabel] at (-12, 3)
  {Antenna\\R\&S\textsuperscript{\textregistered}HK116E};

\draw[box] (2, -3) rectangle (22, 9);
\node[boxlabel] at (12, 3)
  {Antenna\\MVG QH2000};


\draw[box] (-13, -30) rectangle (13, -18);
\node[boxlabel] at (0, -24)
  {Signal Analyzer\\R\&S\textsuperscript{\textregistered}FSW43};


\draw[box] (-13, -50) rectangle (13, -38);
\node[boxlabel] at (0, -44)
  {Fraunhofer HHI\\Synchronomat};


%
%
\draw (-12, -4) -- (-12, -12);                 
\draw ( 12, -4) -- ( 12, -12);                 
\draw (-12,-12) -- ( 12,-12);                  
\draw[conn] (0, -12) -- (0, -17);              

\node[arrlabel, anchor=east]            at (-13, -8) {300\,MHz};
\node[arrlabel, anchor=west, align=left] at ( 13, -8) {5\,GHz\\27\,GHz};

%
%
\draw[conn] (-2, -37) -- (-2, -31);            
\draw[conn] ( 2, -37) -- ( 2, -31);            
\node[arrlabel, anchor=east] at (-3, -34) {10\,MHz};
\node[arrlabel, anchor=west] at ( 3, -34) {Trigger};

\end{tikzpicture}}
    \caption{}
    \label{fig:block_rx}
  \end{subfigure}
  \caption{Block diagram of the (a) transmitter and (b) receiver measurement chains.}
  \label{fig:blockdiagram}
\end{figure}
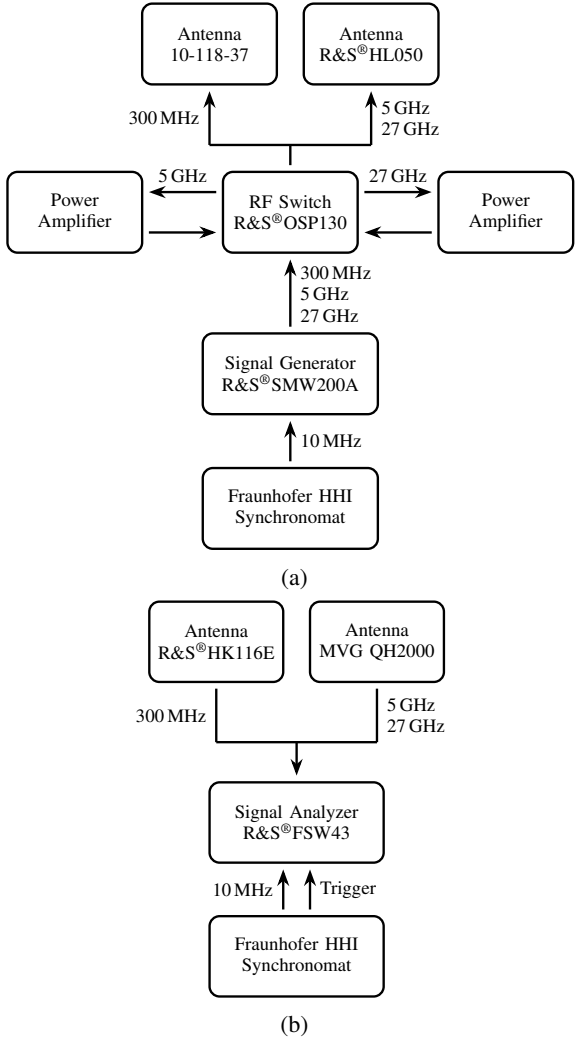

The channel sounder was composed of commercial test and measurement equipment; Fig.~\ref{fig:blockdiagram} shows the block diagram of the measurement chains. A vector signal generator R\&S\textregistered SMW200A was used at the transmitter side, and a signal and spectrum analyzer R\&S\textregistered FSW was utilized for the sampling of the received signals at the receiver side. The different frequencies were measured one after another, with antennas, amplifiers, waveforms, and carrier frequencies switched automatically. Vertically polarized antennas were used at all three carrier frequencies. Tx and Rx were synchronized in frequency and time by two Rubidium clocks (HHI Synchronomat) that had been aligned in advance. Furthermore, back-to-back calibration measurements of Tx and Rx were performed for each measurement frequency: Tx and Rx were connected directly via cable, capturing the frequency response of the sounder hardware, which was subsequently removed from all measurements in the frequency domain. This normalization also removes the impairments of the other measurement equipment.

Table~\ref{tab:cs_parameters} shows the parameters of the channel sounder, which were selected and optimized with respect to the expected impact of the rotors on the wave propagation according to Table~\ref{tab:helicopter_parameters}. The measurement bandwidths were constrained by regulatory issues. Although a broader bandwidth is favorable in general -- because it enables a higher delay resolution -- it is of minor importance for this study, which focuses on the time evolution and on the Doppler frequencies of the direct LOS path. Here, a high temporal resolution, a high maximum Doppler frequency, and a long measurement time are of importance.
\begin{table}[htbp]
\caption{Parameters of the channel sounder}
\label{tab:cs_parameters}
    \begin{center}
        \begin{NiceTabular}[c]{l l l l}
        \CodeBefore
        \rowcolor{gray!50}{1-1}
        \rowcolors{2}{gray!15}{}
        \Body
            \toprule
			\textit{Parameter} & \textit{UHF} & \textit{C-band} & \textit{Ka-band}\\
			\midrule
            Carrier frequency & 302 MHz & 4.9 GHz & 27.1 GHz\\
            Measurement bandwidth & 5 MHz & 500 MHz & 25 MHz\\
            Sequence length ($M = N$) & 100 & 10,000 & 250\\
            No. of sequences ($K = L$) & 20,000 & 20,000 & 40,000\\
            Seq. duration / temp. resolution & 20 µs & 20 µs & 10 µs\\
            Measurement time  & 400 ms & 400 ms & 400 ms\\
            Delay resolution  & 40 ns & 2 ns & 20 ns\\
            Doppler resolution  & 1.25 Hz & 1.25 Hz & 1.25 Hz \\
            Frequency resolution  & 50 kHz & 50 kHz & 100 kHz\\
            Maximum Doppler freq. & 25 kHz & 25 kHz & 50 kHz\\
            Antenna gain Rx & $\sim$3 dBi & 7 dBi & 15 dBi\\
            Antenna gain Tx & $\sim$3 dBi & 8.5 dBi & 8.5 dBi\\
			\bottomrule
        \end{NiceTabular}
    \end{center}
\end{table}
\section{Evaluation of Measurements}
The measurement data were captured by the signal analyzer and afterward processed using MATLAB\textregistered. The processing included the application of the calibration to the raw data, yielding the Input Delay Spread Function. Depending on the parameters of interest, different profiles were evaluated with appropriate window functions and oversampling applied. The evaluation parameters for the time-variant analyses are summarized in Table~\ref{tab:param_eval}. The window functions were used to filter unwanted parts of the system functions in the delay and Doppler frequency domains and to improve the signal-to-noise ratio in the conjugate dimensions of frequency and time. Furthermore, windowing in the frequency and time domains was used to suppress sidelobes in the conjugate domains, i.e., delay and Doppler frequency. In the first case, a flat-top window such as the Tukey window is applied, whereas in the latter case a window with good sidelobe suppression such as the ultra-spherical window with parameter $\lambda = 1$ is used \cite{doerry}.
\begin{table}[htbp]
\caption{Evaluation parameters for the time-variant analyses}
\label{tab:param_eval}
    \begin{center}
        \begin{NiceTabular}[c]{l l}
        \CodeBefore
        \rowcolor{gray!50}{1-1}
        \rowcolors{2}{gray!15}{}
        \Body
            \toprule
			 \textit{Parameter} & \textit{Value}\\
			\midrule
   \multicolumn{2}{l}{\textit{Common (DPS, spectrogram, PTP)}}\\
   Filter delay domain, 302 MHz & Tukey, \sfrac{1}{2}, 500 ns\\
   Filter delay domain, 4.9 GHz & Tukey, \sfrac{1}{2}, 5 ns\\
   Filter delay domain, 27.1 GHz & Tukey, \sfrac{1}{2}, 100 ns\\
   Delay compensation & yes\\
   \midrule
   \multicolumn{2}{l}{\textit{Doppler Power Spectrum}}\\
   Oversampling Doppler domain & 4 \\
   Filter time domain & ultra-spherical, 70 dB \\
   Eval. threshold for RMS Doppler spread & 50 dB\\
   \midrule
   \multicolumn{2}{l}{\textit{Doppler spectrogram}}\\
   Filter Doppler domain, 302 MHz & Tukey, \sfrac{1}{2}, 2 kHz \\
   Filter Doppler domain, 4.9 GHz & Tukey, \sfrac{1}{2}, 3 kHz \\
   Filter Doppler domain, 27.1 GHz & Tukey, \sfrac{1}{2}, 6 kHz \\
   Window length (302 MHz, 4.9 GHz) & 10 ms\\
   Window length (27.1 GHz) & 5 ms\\
   Overlap & \sfrac{1}{2} \\
   Window & Kaiser, $\beta$ = 4 \\
			\bottomrule
        \end{NiceTabular}
    \end{center}
\end{table}
\subsection{Static Measurements: Instantaneous Power Delay Profile}
\begin{figure}[htbp]
  \centering
  \includegraphics[scale = 0.88]{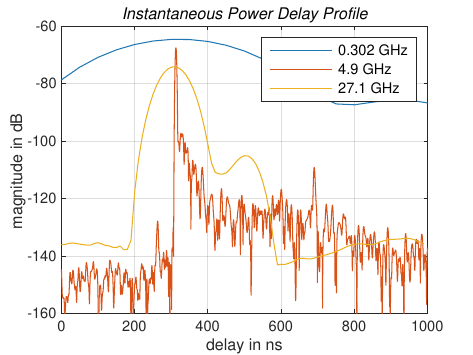}
  \caption{Instantaneous power delay profile for the Tiger (static measurement).}
  \label{fig:PDP_tiger}
\end{figure}

Measurements with stationary rotor blades were performed to verify the measurement setup. For this purpose, the Input Delay Spread Function was filtered with an ultra-spherical window (70~dB sidelobe suppression) in the frequency domain, oversampled by a factor of four in the delay domain, and coherently averaged over the time domain, yielding the instantaneous power delay profile (IPDP). Fig.~\ref{fig:PDP_tiger} shows the result for the Tiger at all three frequencies as a representative example. The measurement at 4.9~GHz demonstrates an exceptionally high dynamic range of more than 70~dB, a strong LOS component, and some weak multipath components at excess delays of up to a few hundred nanoseconds, which stem from the specific measurement environment and are not of interest here; the delay resolution at the other two frequencies is lower due to the much smaller measurement bandwidths. The maximum of the IPDP corresponds to the free-space path loss, which differs between the three measurements due to the different carrier frequencies and antenna gains (see Table~\ref{tab:cs_parameters}). In the following, the evaluations concentrate on the LOS path: the delay is estimated from the linear phase of the Time-Variant Transfer Function and compensated for (Delay compensation, see Table~\ref{tab:param_eval}), and the Input Delay Spread Function is subsequently windowed in the delay domain to remove all multipath components.
\subsection{Time-Variant Measurements: Doppler Power Spectrum}
\begin{figure}[htbp]
  \centering
  \includegraphics[scale = 0.88]{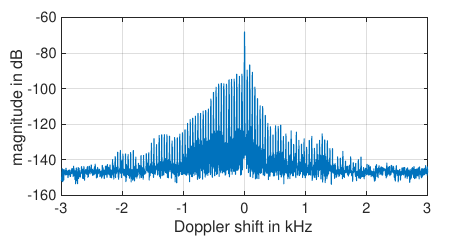}
  \includegraphics[scale = 0.88]{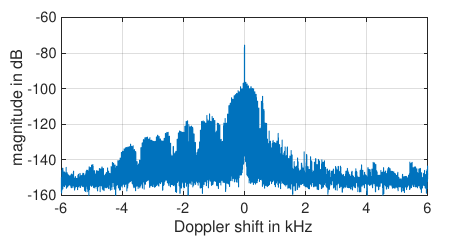}
  \caption{DPS of the UMAT, 4.9 GHz (top), 27.1 GHz (bottom).}
  \label{fig:DPS_UMAT}
\end{figure}

\begin{figure}[htbp]
  \centering
  \includegraphics[scale = 0.88]{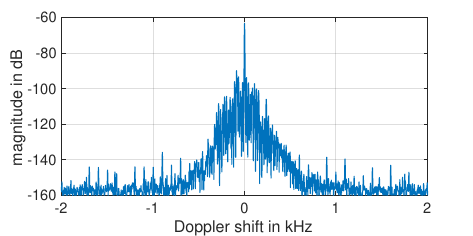}
  \includegraphics[scale = 0.88]{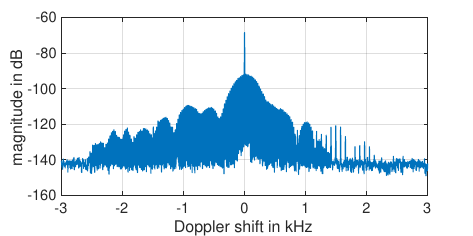}
  \includegraphics[scale = 0.88]{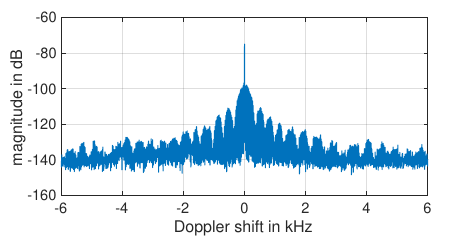}
  \caption{DPS of the Tiger, 302 MHz (top), 4.9 GHz (center), 27.1 GHz (bottom).}
  \label{fig:DPS_Tiger}
\end{figure}

\begin{figure}[htbp]
  \centering
  \includegraphics[scale = 0.88]{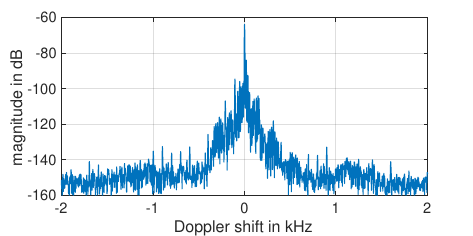}
  \includegraphics[scale = 0.88]{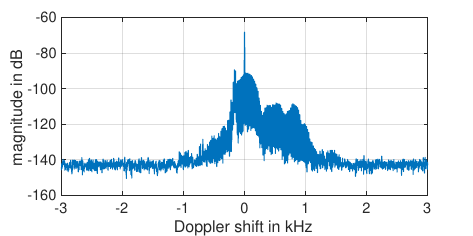}
  \includegraphics[scale = 0.88]{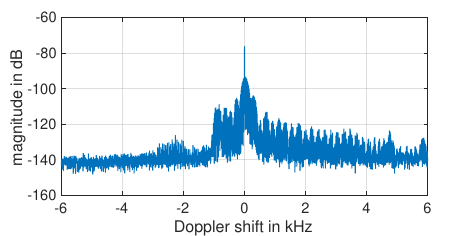}
  \caption{DPS of the CH-53, 302 MHz (top), 4.9 GHz (center), 27.1 GHz (bottom).}
  \label{fig:DPS_CH53}
\end{figure}

In the case of dynamic measurements with spinning rotor blades, the delay-compensated and delay-windowed (to isolate the LOS path) Input Delay Spread Functions were transformed to the Delay-Doppler Spread Function and summed over the delay domain to yield the Doppler Power Spectrum (DPS). Additionally, a window in the time domain was applied to suppress the sidelobes in the Doppler domain; Table~\ref{tab:param_eval} shows the applied evaluation parameters. Figs.~\ref{fig:DPS_UMAT}, \ref{fig:DPS_Tiger}, and \ref{fig:DPS_CH53} show the DPS of the UMAT, the Tiger, and the CH-53 at all measured carrier frequencies (no measurements were performed for the UMAT at 302~MHz). Due to the periodic nature of the time variance, the DPS is a line spectrum. The line spacing equals the blade-passage frequency $f_\mathrm{bp} = N_\mathrm{b} \, f_\mathrm{rot}$ (see Table~\ref{tab:helicopter_parameters}), i.e., 42.5~Hz for the UMAT, 21.8~Hz for the Tiger, and 18.5~Hz for the CH-53.

Table~\ref{tab:RMS_DPS} shows the RMS Doppler spread of all helicopters and frequencies, which quantifies the expected Doppler shifts due to the time variance and is defined as the square root of the second central moment of the DPS \cite{parsons}, \cite{keusgen}:
\begin{equation}
\nu_\mathrm{rms} = \sqrt{\int \left(\nu - \overline{\nu}\right)^2 S_D(\nu)\,\mathrm{d}\nu \bigg/ \int S_D(\nu)\,\mathrm{d}\nu}\,,
\label{eq:rms_doppler}
\end{equation}
where $S_D(\nu)$ denotes the DPS and $\overline{\nu}$ its center of gravity. Spectral components more than 50~dB below the DPS peak are excluded from the integrals (see Table~\ref{tab:param_eval}) to prevent noise from inflating the spread estimate. The estimated Doppler spread values are relatively small, despite the low evaluation threshold. This can be attributed to the predominantly lateral movement of the blades relative to the LOS path. At 302~MHz, the spectra are confined to a few hundred hertz around the carrier, consistent with the roughly 16-fold and 90-fold smaller Doppler shifts compared to 4.9~GHz and 27.1~GHz, respectively.
\begin{table}[htbp]
\caption{RMS Doppler spreads}
\label{tab:RMS_DPS}
    \begin{center}
        \begin{NiceTabular}[c]{r r r r}
        \CodeBefore
        \rowcolor{gray!50}{1-1}
        \rowcolors{2}{gray!15}{}
        \Body
            \toprule
			 & \textit{UMAT} & \textit{Tiger} & \textit{CH-53}\\
			\midrule
   302 MHz & -- & 8.6 Hz & 5.6 Hz\\
   4.9 GHz & 39.6 Hz & 46.0 Hz & 39.3 Hz\\
   27.1 GHz & 75.5 Hz & 58.9 Hz & 105.8 Hz\\
			\bottomrule
        \end{NiceTabular}
    \end{center}
\end{table}
\subsection{Time-Variant Measurements: Doppler Spectrogram}
The DPS describes the distribution of the Doppler frequencies within the measurement interval but gives no information about their temporal evolution. This information can be retrieved by a temporal-spectral analysis \cite{boashash}, \cite{chen} applied to a time series of samples. Therefore, we propose to reduce the Input Delay Spread Function to a series of zero-bandwidth channel impulse responses by complex summation along the delay axis, in contrast to the summation of powers used for the Power Time Profile (see Fig.~\ref{fig:bello}) \cite{keusgen}. The complex summation over the delay axis equals the Time-Variant Transfer Function evaluated at the center frequency, $T(t,0)$; since the Input Delay Spread Function was delay-compensated and windowed to the LOS path beforehand, this coherent summation thus yields a single complex channel-gain time series that corresponds to a narrowband observation of the LOS channel. To this time series, we apply the spectrogram as an estimator for the temporal-spectral analysis.

\begin{figure*}[!t]
  \begin{center}
    \begin{subfigure}[t]{5.9cm}
    \centering
      \parbox[b][3.554cm][c]{5.9cm}{\centering\footnotesize no measurement\\at 302 MHz}\\
      \includegraphics[width=5.9cm]{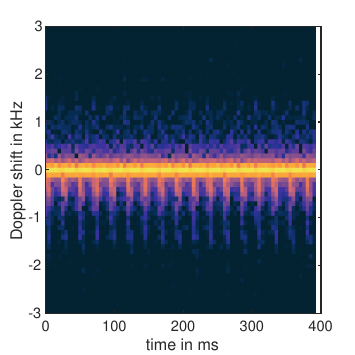}\\
      \includegraphics[width=5.9cm]{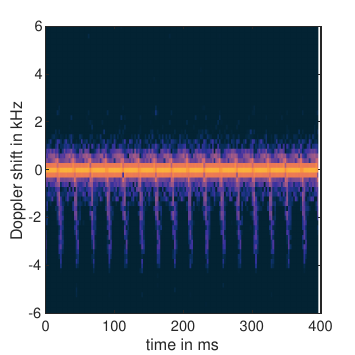}
      \caption{UMAT, 4.9 GHz (center), 27.1 GHz (bottom)}
      \label{fig:spec_umat}
    \end{subfigure}
    \vline\hfill
    \begin{subfigure}[t]{5.9cm}
    \centering
      \begin{tikzpicture}
        \node[anchor=south west, inner sep=0] (img) {\includegraphics[width=5.9cm]{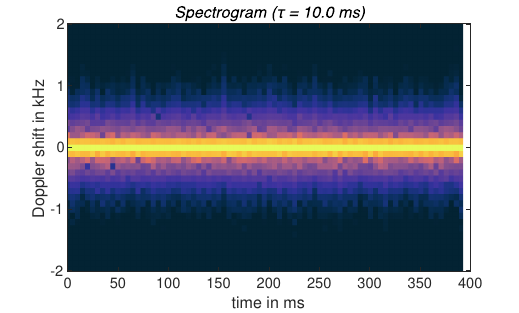}};
        \fill[white] ([yshift=-0.22cm]img.north west) rectangle (img.north east);
      \end{tikzpicture}\\
      \includegraphics[width=5.9cm]{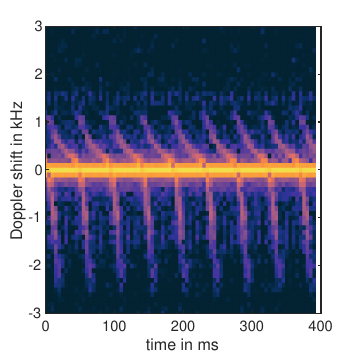}\\
      \includegraphics[width=5.9cm]{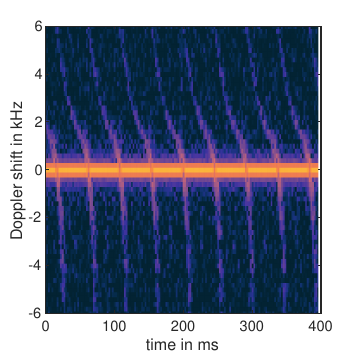}
      \caption{Tiger, 302 MHz (top), 4.9 GHz, 27.1 GHz}
      \label{fig:spec_tiger}
    \end{subfigure}
    \vline\hfill
    \begin{subfigure}[t]{5.9cm}
    \centering
      \begin{tikzpicture}
        \node[anchor=south west, inner sep=0] (img) {\includegraphics[width=5.9cm]{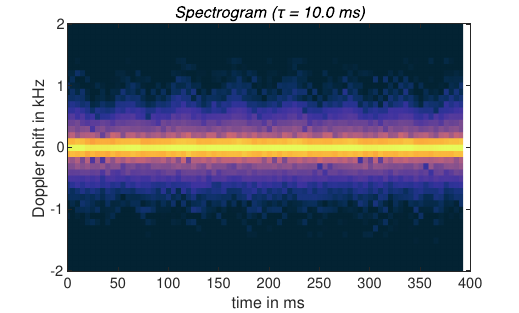}};
        \fill[white] ([yshift=-0.22cm]img.north west) rectangle (img.north east);
      \end{tikzpicture}\\
      \includegraphics[width=5.9cm]{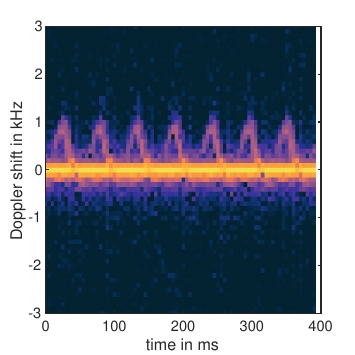}\\
      \includegraphics[width=5.9cm]{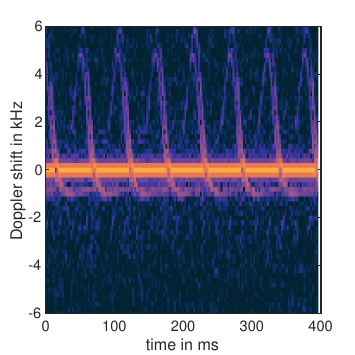}
      \caption{CH-53, 302 MHz (top), 4.9 GHz, 27.1 GHz}
      \label{fig:spec_ch53}
    \end{subfigure}
  \end{center}
  \caption{Doppler spectrograms of all helicopters; helicopters are columns, carrier frequencies are rows (302 MHz, 4.9 GHz, 27.1 GHz from top to bottom); power in dB ranging from $-125$~dB (black) to $-65$~dB (yellow).}
  \label{fig:spec}
\end{figure*}

Fig.~\ref{fig:spec} shows the spectrograms of all helicopters at all measured carrier frequencies. At 302~MHz (Tiger and CH-53), the spectrograms are almost featureless: the energy is concentrated in a narrow band around the LOS line, and only weak periodic variations are visible -- itself a noteworthy result, as it confirms that the rotor influence on the Doppler domain is negligible at UHF. At 4.9~GHz and 27.1~GHz, the periodic structure arising from the blade-to-blade time (see Table~\ref{tab:helicopter_parameters}) is clearly visible, as are specific temporal signatures, which are resolved differently depending on the carrier frequency. These signatures repeat with the blade-to-blade period and become sharper with increasing carrier frequency.
\subsection{Time-Variant Measurements: Power Time Profile}
\begin{figure}[htbp]
  \centering
  \includegraphics[scale = 0.88]{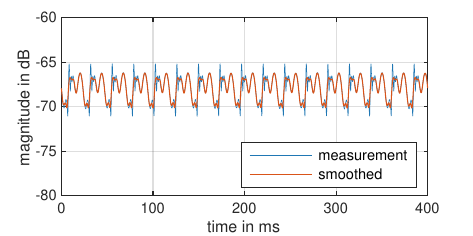}
  \includegraphics[scale = 0.88]{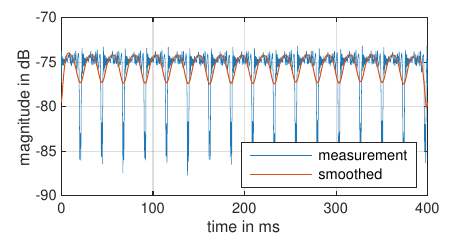}
  \caption{PTP of the UMAT, 4.9 GHz (top), 27.1 GHz (bottom); measured (blue) and smoothed (red).}
  \label{fig:PTP_UMAT}
\end{figure}

\begin{figure}[htbp]
  \centering
  \includegraphics[scale = 0.88]{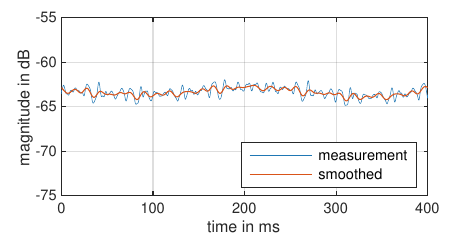}
  \includegraphics[scale = 0.88]{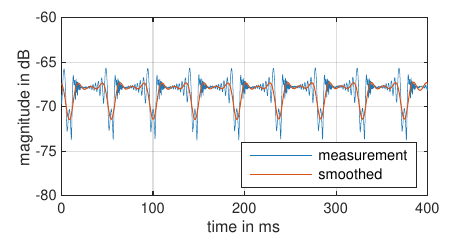}
  \includegraphics[scale = 0.88]{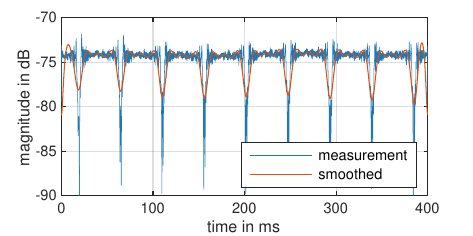}
  \caption{PTP of the Tiger, 302 MHz (top), 4.9 GHz (center), 27.1 GHz (bottom); measured (blue) and smoothed (red).}
  \label{fig:PTP_Tiger}
\end{figure}

\begin{figure}[htbp]
  \centering
  \includegraphics[scale = 0.88]{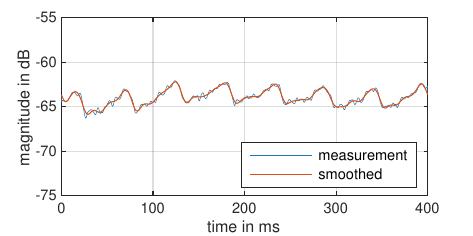}
  \includegraphics[scale = 0.88]{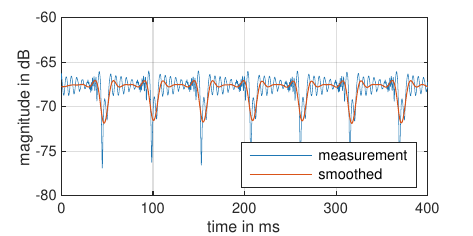}
  \includegraphics[scale = 0.88]{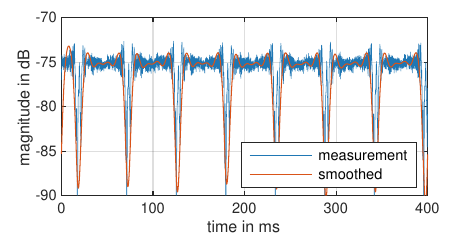}
  \caption{PTP of the CH-53, 302 MHz (top), 4.9 GHz (center), 27.1 GHz (bottom); measured (blue) and smoothed (red).}
  \label{fig:PTP_CH53}
\end{figure}

Besides the Doppler frequencies, the temporal modulation of the channel gain, i.e., of the received power, is of specific interest. It is characterized by the Power Time Profile (PTP), which is obtained from the appropriately filtered Input Delay Spread Function by summation of the powers along the delay axis (see Fig.~\ref{fig:bello} and Table~\ref{tab:param_eval}). Figs.~\ref{fig:PTP_UMAT}, \ref{fig:PTP_Tiger}, and \ref{fig:PTP_CH53} depict the resulting PTPs of all three helicopters in blue color. The periodic power modulation at the blade-to-blade period (see Table~\ref{tab:helicopter_parameters}) is clearly visible and exactly matched. Furthermore, the PTP exhibits a high variation (ripple), especially within the blockage events; the effect is present at all frequencies and more pronounced at the higher carrier frequencies.

To allow for a quantitative analysis of the power loss due to blockage, it is useful to smooth this ripple. Here, we propose to perform the smoothing by lowpass filtering in the Doppler domain, where the cut-off frequency is derived from the width $T_B$ of the blockage event as $\nu_\mathrm{cutoff} = \sfrac{1}{2}\cdot\sfrac{1}{T_B}$. A Tukey window was applied, with the cut-off frequency defined as the width of its flat top. The approximate blockage durations read from the measured PTPs are 3.5~ms (UMAT), 6~ms (Tiger), and 9~ms (CH-53), resulting in cut-off frequencies of 143~Hz, 84~Hz, and 55.5~Hz, respectively. The rationale for this choice of cut-off frequency is that spectral components above $1/(2 T_B)$ mainly represent the diffraction ripple within a blockage event, whereas the envelope of the blockage itself is confined to frequencies below this value; the filtering therefore removes the ripple while preserving the depth and the duration of the blockage. The measured blockage durations are of the same order as the state-duration statistics reported for the Ku-band satellite-to-helicopter channel, where mean blade-obstruction durations of 5.4~ms and 12~ms were found for the Hughes H-500 and the Seahawk SH-60B, respectively \cite{lemoscid2}. Figs.~\ref{fig:PTP_Tiger} and \ref{fig:PTP_CH53} additionally show the smoothed PTPs (in red color), which were derived after reducing the Doppler bandwidth according to these cut-off frequencies.

\begin{figure}[htbp]
  \centering
  \resizebox{0.88\columnwidth}{!}{\begin{tikzpicture}
\begin{axis}[
  scale only axis=true,
  width=216pt,
  height=113pt,
  axis lines*=left,                       
  axis line style={line width=0.6pt},     
  tick style={line width=0.6pt},          
  axis on top=false,                      
  tick align=outside,                     
  %
  ymin=0, ymax=17,
  xmin=-0.5, xmax=2.5,
  enlarge x limits=false,
  %
  ylabel={Blockage attenuation in dB},
  xlabel={Carrier frequency},
  xtick={0,1,2},
  xticklabels={302\,MHz, 4.9\,GHz, 27.1\,GHz},
  ytick={0,3,6,9,12,15},
  every axis label/.append style={font=\footnotesize},
  every tick label/.append style={font=\footnotesize},
  %
  ymajorgrids=true,
  xmajorgrids=false,
  grid style={dotted, gray!60, line width=0.4pt},
  %
  ybar,
  bar width=0.25,
  %
  nodes near coords,
  nodes near coords style={
    font=\scriptsize, inner sep=1pt,
    /pgf/number format/.cd, fixed, fixed zerofill, precision=1,
  },
  %
  legend pos=north west,
  legend style={
    draw=none, fill=none,
    font=\footnotesize,
    inner sep=1pt, row sep=-1pt,
  },
  legend cell align=left,
  legend image code/.code={
    \draw[#1, draw=black, line width=0.5pt]
      (0cm,-0.7mm) rectangle (3mm,1.4mm);
  },
]

  \addplot[fill=cmapBlue, draw=black, line width=0.6pt, bar shift=-0.25]
    coordinates {(1,3.0) (2,9.1)};
  \addlegendentry{UMAT}

  \addplot[fill=cmapOrange, draw=black, line width=0.6pt, bar shift=0]
    coordinates {(1,4.7) (2,6.4)};
  \addlegendentry{Tiger}

  \addplot[fill=cmapGreen, draw=black, line width=0.6pt, bar shift=0.25]
    coordinates {(0,2.3) (1,4.0) (2,15.0)};
  \addlegendentry{CH-53}

  \draw[draw=cmapGray, line width=0.5pt, fill=white,
        postaction={pattern=north east lines,
                    pattern color=cmapGray}]
    (axis cs:-0.375,0) rectangle (axis cs:-0.125,1.0);
  \draw[draw=cmapGray, line width=0.5pt, fill=white,
        postaction={pattern=north east lines,
                    pattern color=cmapGray}]
    (axis cs:-0.125,0) rectangle (axis cs:0.125,1.0);

\end{axis}
\end{tikzpicture}}
  \caption{Estimated attenuation caused by blade blockage (hatched: no measurement at 302 MHz for the UMAT; no reliable estimate for the Tiger).}
  \label{fig:att_bar}
\end{figure}
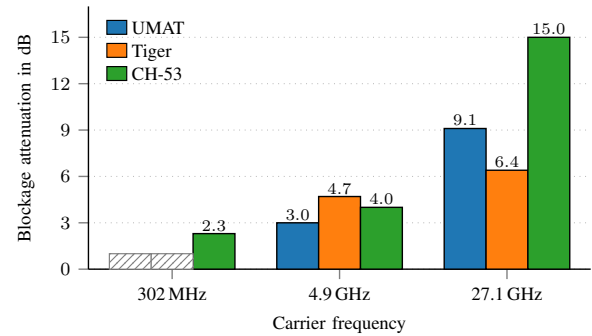

Finally, Fig.~\ref{fig:att_bar} shows the estimated blockage attenuation for all helicopters and frequencies, obtained from the smoothed curves by comparing the minimum value within a blockage event with the minimum value between two blockage events. At 302~MHz, the power modulation was very small (see the top rows of Figs.~\ref{fig:PTP_Tiger} and \ref{fig:PTP_CH53}): for the Tiger, the ripple was too indistinct to extract a reliable attenuation value, whereas for the CH-53 a clear periodic modulation at the blade-to-blade period remains visible and a blockage attenuation of 2.3~dB was estimated. The blockage attenuation increases with the carrier frequency, which is qualitatively consistent with diffraction theory: for a fixed blade geometry and measurement distance, the blade width measured in Fresnel zones grows with frequency, deepening the diffraction shadow. The amount of the increase is specific to the rotor blade; no clear correlation between the blade width and the attenuation increase is observed (the UMAT has the smallest blades but not the smallest attenuation at 27.1~GHz).
\subsection{Time-Variant Measurements: Modeling of the Power Time Profile}
In the following, the modeling of the blockage attenuation by means of the double knife-edge (DKE) model \cite{kunisch} was investigated. This model -- which represents the obstacle by a thin conducting strip, i.e., by two parallel knife edges -- is expected to fit the measured attenuation well, as the rotor blades closely resemble a metallic strip in shape. The DKE model treats the rotating blade quasi-stationarily as a sequence of stationary strip positions. A rigorous analytical treatment of the rotating strip based on Hertzian electrodynamics and the Wiener--Hopf technique \cite{dasbasi} critiques quasi-stationary approaches on causality grounds and predicts diffracted-field components at harmonics of the rotation; for the blade-passage-periodic quantities considered here, the quasi-stationary DKE model nevertheless proves sufficiently accurate, as shown below.

\begin{figure}[htbp]
  \centering
  \includegraphics[scale = 0.88]{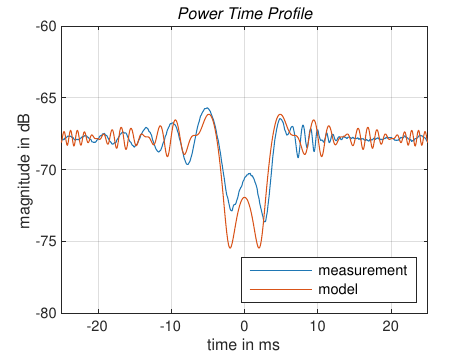}
  \includegraphics[scale = 0.88]{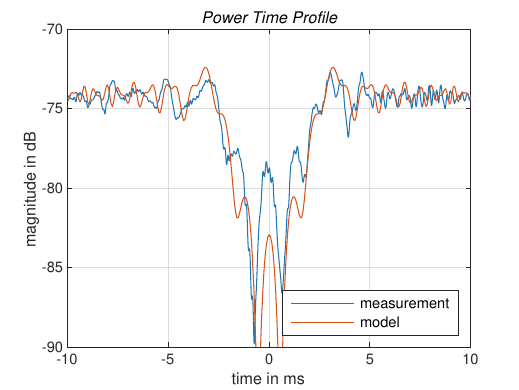}
  \caption{Measured attenuation (blue) and double knife-edge model (red) for the Tiger at 4.9 GHz (top) and 27.1 GHz (bottom).}
  \label{fig:DKE_Tiger}
\end{figure}

Since the basic geometry as well as the width and the velocity of the rotor blades are known, the model from \cite{kunisch} could be parameterized directly: for the Tiger, the distance from the Tx to the blade intercept point is $d_1 \approx 8.8$~m (see Table~\ref{tab:geometry}), the distance from the intercept point to the Rx is $d_2 \approx 85$~m, the blade width is $w = 0.5$~m, and the blade velocity at the intercept point is $v \approx 156$~\sfrac{m}{s} (70\% of the tip speed). Fig.~\ref{fig:DKE_Tiger} compares the modeled blockage attenuation (red) with one typical measured blockage event (blue, centered around time zero) for the Tiger at 4.9~GHz and at 27.1~GHz. Good agreement is observed between the measurement and the model at both carrier frequencies, in the depth and the duration of the blockage event as well as in the characteristic double-dip shape and the adjacent diffraction ripple, which are clearly resolved at 4.9~GHz.

It should be noted that the DKE model of \cite{kunisch} was originally derived for the obstruction of a ray by a person at low elevation angles and for frequencies between 4 and 10~GHz. Its application at elevation angles of 25\textdegree{}--30\textdegree{}, at carrier frequencies up to 27.1~GHz, and to a metallic obstacle therefore constitutes an extrapolation, which is, however, well supported by the agreement observed in Fig.~\ref{fig:DKE_Tiger}.
\section{Conclusion}
This paper has presented an experimental characterization of the rotor blade modulation of three military helicopters at three carrier frequencies between 302~MHz and 27.1~GHz. The effect of the rotor blades was characterized in terms of the Doppler Power Spectrum and the RMS Doppler spread, which showed only a moderate impact relative to static channels, with RMS values ranging from 5.6~Hz (CH-53 at 302~MHz) to 105.8~Hz (CH-53 at 27.1~GHz); this is attributed to the predominantly lateral blade motion relative to the LOS path. Additionally, the temporal evolution of the Doppler frequencies (micro-Doppler) was estimated by means of spectrograms and showed distinct helicopter-specific signatures at the blade-passage period, which may be exploited in bistatic or passive radar applications based on rotor blade scattering \cite{clemente}. Furthermore, the power modulation caused by the rotor blade movement was characterized by the Power Time Profile: periodic blockage events were observed, with attenuations increasing from approximately 2--5~dB at UHF and C-band to up to 15~dB at Ka-band and with a period equal to the blade-to-blade time. Finally, the power variation was modeled by means of the double knife-edge model, which shows good agreement with the measured blockage attenuation at both C-band and Ka-band -- including the characteristic double-dip shape of the blockage events -- and thus allows the expected blockage depth to be predicted from the geometry alone. Future work will extend the characterization to the tail rotors, to additional measurement geometries and elevation angles, and to the integration of the measured channel statistics into channel models and link-level simulations for future air-to-ground, air-to-air, and air-to-satellite communication systems.

\end{document}